# Time-Of-Flight methodologies with large-area diamond detectors for the effectively characterization of tens MeV protons

M. Salvadori[a,b,c], P. L. Andreoli[a], M. Cipriani[a], G. Cristofari[a], R. De Angelis[a], S. Malko[d,e], L. Volpe[d,f], J. A. Perez Hernandez[d], J. I. Apiñaniz[d], A. Morace[g], P. Antici[b], M. Migliorati[c,h], G. Di Giorgio[a], F. Consoli[a].

[a] *ENEA, Fusion and Nuclear Safety Department, C. R. Frascati Via E. Fermi 45, Frascati, Italy*

[b] *INRS-EMT Varennes, Québec, Canada*

[c] *Università di Roma La Sapienza*
  *Piazzale Aldo Moro 5, Rome, Italy*

[d] *Centro de Laseres Pulsados (CLPU) Parque Cientifico, E-37185 Villamayor, Salamanca, Spain*

[e] *Princeton Plasma Phyics Laboratory 100 Stellarator Rd. Princeton, New Jersey 08536, United States*

[f] *Laser-Plasma Chair at the University of Salamanca, Salamanca, Spain*

[g] *Institute of Laser Engineering Osaka University, Japan*

[h] *INFN sez. Roma Piazzale Aldo Moro 2, Roma, Italy*

*E-mail*: martina.salvadori@uniroma1.it

ABSTRACT: A novel detector based on a polycrystalline diamond sensor is here employed in an advanced Time-Of-Flight scheme for the characterization of energetic ions accelerated during laser-matter interactions. The optimization of the detector and of the advanced TOF methodology allow to obtain signals characterized by high signal-to-noise ratio and high dynamic range even in the most challenging experimental environments, where the interaction of high-intensity laser pulses with matter leads to effective ion acceleration, but also to the generation of strong Electromagnetic Pulses (EMPs) with intensities up to the MV/m order. These are known to be a serious threat for the fielded diagnostic systems. In this paper we report on the measurement performed with the PW-class laser system Vega 3 at CLPU (∼ 30 J energy, ∼ $10^{21}$ W/cm$^2$ intensity, ~30 fs pulses) irradiating solid targets, where both tens of MeV ions and intense EMP fields were generated. The data were analyzed to retrieve a calibrated proton spectrum and in particular we focus on the analysis of the most energetic portion (E > 5.8 MeV) of the spectrum showing a procedure to deal with the intrinsic lower sensitivity of the detector in the mentioned spectral-range.

# Contents

## 1. Introduction

During the past twenty years the progresses in laser-driven acceleration mechanisms led to the proposition to use them as an alternative source of energetic particles for many different applications [1,2]; jointly, the maximum proton energy achievable by laser-matter interactions saw a significant growth. This was possible thanks to the larger availability of powerful laser systems that allowed several groups to work independently on the topic. This scenario was a fertile soil for the exploitation of new-target designs and hybrid acceleration schemes trying to push the proton cutoff energies to higher values [3–6].
The ions accelerated by these processes typically present a broadband spectrum with the number of particles spanning over several order of magnitudes and decreasing for increasing energy [1].
     The complete and accurate characterization of the whole ion spectrum is therefore a challenging task. Indeed, it requires to have a diagnostic system characterized by high sensitivity to appreciate the maximum achievable energy, but also to have a high dynamic range.
     Moreover, the effective employment of laser-driven ion sources in various applications depends on the possibility to work at high-repetition rates. From the laser-side, the technology is already mature to guarantee repetition rate of tens of Hz while maintaining high energy and high intensities on target with good stability. The targetry is also moving towards the development of solutions that would allow to run almost continuously, for instance by using gas jet targets [7], cryogenic jets [8], liquid crystal films [9] and liquid targets [10] that are currently under investigation. It is essential that also the diagnostic system used to monitor the interaction can work at the same rate. This can be a challenging task. Indeed, when a high intensity laser interacts with matter, intense electromagnetic waves in the microwave-radiofrequency range are also produced [11]. These are proven to be a serious threat for any electronic device placed near the interaction point, leading to the disabling, or even to the damaging, of the deployed diagnostic systems. The use of passive detectors generally solves the problem, but this is not a viable solution when aiming for an on-line characterization of a system working at high repetition rates. The resistance to the presence of strong electromagnetic fields is therefore an essential requirement to fulfil.
To meet the high sensitivity and dynamic range requirements while granting a high EMP resistance, we developed a detector based on a polycrystalline diamond sensor having large area (15 mm × 15 mm) and 150 µm thickness equipped with an optimized shielding structure specifically designed to prevent the detrimental effect due to the coupling of the detector with the electromagnetic waves produced during the interaction [12]. The detector has been off-line characterized by exposition to monochromatic alpha particles having 5.486 MeV energy emitted by a $^{241}$Am radioactive source highlighting a temporal response and a charge collection efficiency (CCE) of 4.1 ns and ~45 % respectively [12]. The detector is meant to be used in Time of Flight (ToF) schemes, a technique adopted within the laser-plasma community to have a prompt characterization of the ion beam parameters with good accuracy [13–17].

In the following sections the employment of the detector for the efficient characterization of tens of MeV protons is discussed. The thickness of 150 µm was indeed chosen as a good trade-off between good temporal resolution and high energy particle sensitivity. The features of the detector were tested during an experimental campaign carried out at the VEGA III laser facility ($E_L$ = 30 J; $\tau_L$ = 30 fs; for a peak power of 1 PW at 1 Hz). The obtained results are here used to highlight the detector performance. During the experimental campaign, the polycrystalline diamond was capable to retrieve proton spectra up to a maximum energy of ~ 20 MeV.

## 2. The polycrystalline diamond detector for ToF

In the time of flight technique, the energy of the particles is retrieved by computing the time they need to travel through a known distance, i.e., from the laser-matter interaction point to the detector (hereinafter $d_{ToF}$). Indeed, during the interaction both photons, electrons and ions are emitted. All of them are going to interact with the sensor used for the measurement and they will produce a signal whose amplitude will depend on the detector and its specific response.

When time of flight technique is performed by means of semiconductor detectors, the acquired signal typically shows a first peak due to the detection of photons ("photopeak") with a descending tail where electrons are contributing and, later, one or more peaks due to the contribution of various populations of accelerated ions [16–18]. The instant of detection of the photopeak, $t_{ph}$, works as an absolute precise time-reference to compute the actual interaction instant $t_{bang} = t_{ph} - \frac{d_{ToF}}{c}$ from which the time-of-flight of the oncoming ions can be retrieved. Then, their energy can be estimated by the following relation:

$$E_i = m_i \, (\gamma_i - 1) \, c^2 \tag{1}$$

where $c$ is the speed of light, $m_i$ is the ion mass, $t_i$ is the ion detection time and

$\gamma_i = \left( 1 - \frac{d_{ToF}^2}{c^2(t_i - t_{bang})^2} \right)^{-1/2}$ the relativistic parameter.

A typical signal collected by the polycrystalline diamond detector is shown in Figure 2.1. This shot was performed irradiating a 3 µm aluminum target with laser energy $E_L$ = 27.9 J, focal spot dimension $\phi$ = 11.67 ± 1.19 µm and temporal duration $\tau_L$ = 323.5 ± 10 fs, for an intensity on target $I_L \simeq 3.5 \times 10^{18}$ Wcm$^{-2}$.

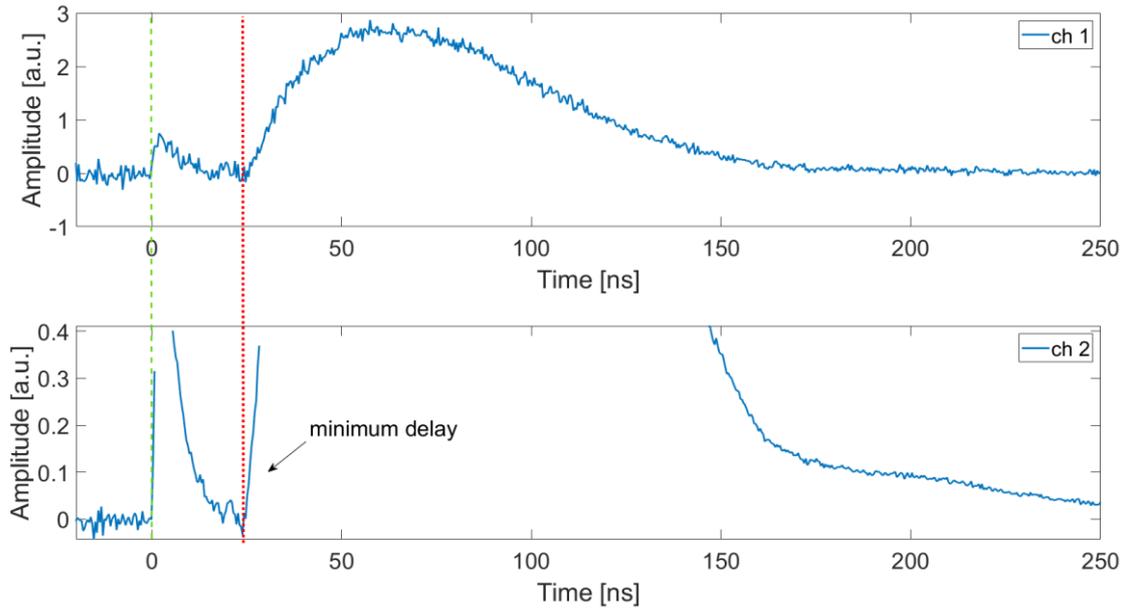

**Figure 2.1** The raw time domain signal collected by the polycrystalline diamond detector during a shot performed on a 3 μm Aluminium foil. The position of the photopeak and the minimum TOF delay are highlighted. Channel 1 provide information on the whole signal whereas the finer details can be retrieved from the signal collected on Channel 2.

This large diamond sensor has high intrinsic sensitivity. To fully exploit this advanced feature of the detector it is important to have a read-out system with large dynamic range. This is a common problem, but the dynamic range can be enhanced by splitting the signal coming from the diamond on two synchronized channels of the same scope (a Tektronix DPO4104B) by means of a calibrated splitter at 50%. The vertical scales of the two channels were set to values which differs for about one order of magnitude. This allowed to record the finer details in one of the channels and to store the information on the whole signal on the other one, as it is shown in Figure 2.1 [16]. This procedure, together with the optimal rejection to EMPs, allowed to retrieve, with great precision, both the time-position of the photopeak and the time instant at which the fastest protons reach the diamond, i.e. the "minimum delay" in the TOF signal (see Figure 2.1 b), from which it is possible to estimate the maximum proton energy for each shot.

Once the energy is known it is possible to infer the number of particles generating the signal by analysing its amplitude. Indeed, when a charged particle enters the bulk of a semiconductor it will generate a certain amount of free electron-hole pairs according to the energy deposited inside the material and to its specific radiation-ionization energy $\epsilon_g$, which is 13.1 eV for diamonds [19]. To produce the signal, these charges have to be collected. Therefore, an external electric field is applied to the semiconductor by means of suitable electrodes. The electrons and holes move apart one from the other trying to reach the electrodes site. Nevertheless, not all of them will successfully traverse the bulk of the detector, indeed some will be lost due to recombination or trapping event. These effects are taken into account by the charge collection efficiency parameter (CCE) that links the number of generated charges, $Q_g$, to those that are effectively collected, $Q_c$. The number of impinging ions can thus be retrieved by applying the relation [16]:

$$N_i = \frac{Q_c \, \epsilon_g}{q_e} \frac{1}{E_i \, CCE} \qquad (2)$$

where $q_e$ is the electronic charge and $Q_c$ is the amount of collected charge that can be estimated by performing a numerical integration of the detected signal $V(t)$, namely:

$$Q_c = k_A \int_{t_i}^{t_f} V(t) dt \qquad (3)$$

With $k_A = \frac{A}{R}$, where $R$ is the impedance of the circuit and $A$ its attenuation.

### 2.1 High energy correction factor

The amount of energy deposited by the particles into the detector depends on their stopping range. It is well known that ions deposit the most of their energy towards the end of their path resulting in the typical Bragg curve describing the energy loss in matter by ions [20].
Therefore, if their range in diamond material is lower than the actual thickness of the detector, they will be completely stopped inside it, releasing all their energy, and equation (2) can be applied as it is. On the other hand, if they would have a longer range, they will cross the whole body of the detector releasing just a portion of their energy in it. In the latter case, the deposited energy does not correspond to the one possessed by the particle itself and estimated thanks to equation (1), thus equation (2) has to be corrected by a proper factor [17, 18].

As mentioned above, the diamond detector has a thickness of 150 µm. According to the simulation performed with the SRIM code [21], within the thickness of the sensor, protons of 5.8 MeV are completely stopped. Up to this energy threshold, protons release all their energy in the bulk of the diamond converting it into e-h pairs according to the relation [16]:

$$Q_g = \frac{q_e E_i}{\epsilon_g} \qquad (4)$$

Knowing the number of e-h pairs produced by the detection of a single particle, it will be possible to retrieve the number of protons impinging onto the detector by analysing the amplitude of the signal, as expressed by equation (2) and (3).

For energies higher than 5.8 MeV, only a portion of the proton total energy will be released inside the detector and equation (4) has to be modified by substituting $E_i$, the energy estimated from the TOF measurements by equation (1), with $E_D$, the portion of the energy lost inside the bulk of the detector by incoming protons with energy $E_i$. This quantity $E_D$ can be determined by means of SRIM simulations. We performed the simulation for proton energies ranging from 1 up to 30 MeV. The results are partially shown in Figure 2.2a where the dE/dx proton energy loss for unit length along the x longitudinal coordinate is depicted for proton energies going from 2 up to 6 MeV. It is possible to see that up to 5 MeV the Bragg peak falls within the 150 $\mu$m of the detector, whereas starting for higher energies only the tail of the curve is visible. The energy $E_D$ deposited inside the detector for each proton energy was thus computed by performing the numerical integration of the obtained curves. Since from the TOF measurements $E_i$ is obtained, it is useful to define a correction factor $R(E_i)$, that links the actual energy deposited inside the diamond to the total energy of the proton estimated by the TOF measurement:

$$E_D = R(E_i) E_i \qquad (5)$$

In Figure 2.2b the obtained correction factor $R(E_i)$ is reported. As expected, it is equal to one for energies lower than 5.8 MeV and then it decreases for higher energy values.

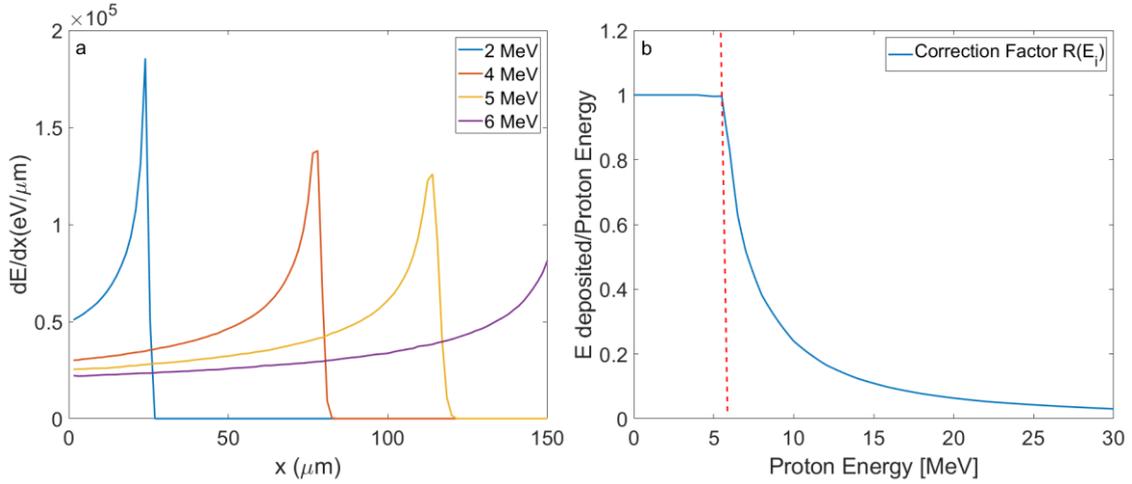

**Figure 2.2 (a)** The energy deposition $\frac{dE}{dx}$ of protons in diamond. the Bragg peak up to 5.8 MeV is visible within the 150 µm thickness of the detector, whereas for higher energies the protons are able to cross the whole detector and escape from the rear side **(b)** The correction factor $R(E_i)$ taking into account that for energies greater that 5.8 MeV, only a portion of the proton total energy is going to be released inside the diamond bulk.

Therefore, equation (2) can be expressed in the more general case as:

$$N_i = \frac{Q_c\, \epsilon_g}{q_e} \frac{1}{R(E_i) E_i\, CCE}, \tag{6}$$

and it thus holds for the whole energy range of protons energies.

## 3. The experimental campaign

The capability of the described diagnostic system to fruitfully characterize high energy ions was tested during an experimental campaign carried out at the VEGA III laser facility at CLPU [22] (Salamanca, Spain). With respect to the TNSA axis, the polycrystalline diamond was placed at 7° horizontal angle and at elevation of 9° of vertical angle. According to the specific needs of energy resolution, the line-of-flight length was adjusted throughout the experiment. Three different lengths were used: $d_{TOF} = 1.498$ m, $d_{TOF} = 1.878$ m and $d_{TOF} = 2.385$ m, resulting in a covered solid angle $d\Omega = 100$ µsr, $d\Omega = 64$ µsr and $d\Omega = 40$ µsr, respectively. A 10 µm aluminium filter was placed in front of the detector surface. This helps to distinguish the ion contribution from the proton component at expenses of the information on protons having energies lower than 0.75 MeV which are completely stopped inside the aluminium filter. The action of the filter also affects the protons with higher energies that are able to reach the detector. Indeed, these are going to reach the diamond sensor with an energy lower than the one estimated by the time-of-flight technique, thus to correctly estimate the number of impinging particles a correction factor, $k_{att}(E_i)$, has to be introduced and equation (6) becomes [16]:

$$N_i = \frac{Q_c\, \epsilon_g}{q_e} \frac{1}{R(E_i) E_i\, k_{att}(E_i) CCE}. \tag{7}$$

The positioning of the aluminium filter in front of the diamond detector contributes also to increase the effectiveness in the shielding action of the diamond case. Indeed, it constitutes the closure of an internal faraday cage providing further protection from the external electromagnetic waves [12]. The diagnostic set up was done following the procedure described in detail in *Salvadori et al.* [16]. This granted an optimal EMP rejection and signal characterized by a high SNR as it highlighted in Figure 2.1.

One of the objectives of the experimental campaign was to exploit the real-time features of this ion diagnostics by collecting a large statistic on the ion bunches accelerated via the TNSA mechanism. Aluminium targets of different thicknesses (ranging from 0.8 µm to 10 µm) were used and for each of them the laser parameters (energy, focal spot dimension and pulse duration) were varied (a paper is in preparation about the full results of the campaign). The real-time detection features of the optimized detection system allowed to collect several consecutive shots. During the experiment, a Thomson spectrometer equipped with a micro channel plate was also deployed. It was placed on the target normal axis at 65 cm from the interaction point. Having the two diagnostic systems placed at similar angles, allowed us to compare the maximum proton energies estimated by the polycrystalline diamond with those provided by the Thomson spectrometer. As it is possible to see in Figure 3.1, a remarkable agreement is achieved, highlighting the capability of the polycrystalline diamond detector to effectively detect tens of MeV protons. These shots were performed by irradiating a 3 µm aluminium target with a laser pulse energy of ~20 J while decreasing the laser pulse duration every few shots. The dashed black lines points out the pulse length variation, the fluctuations in the detected energies within each interval are ascribable to unavoidable small variation in the target alignment.

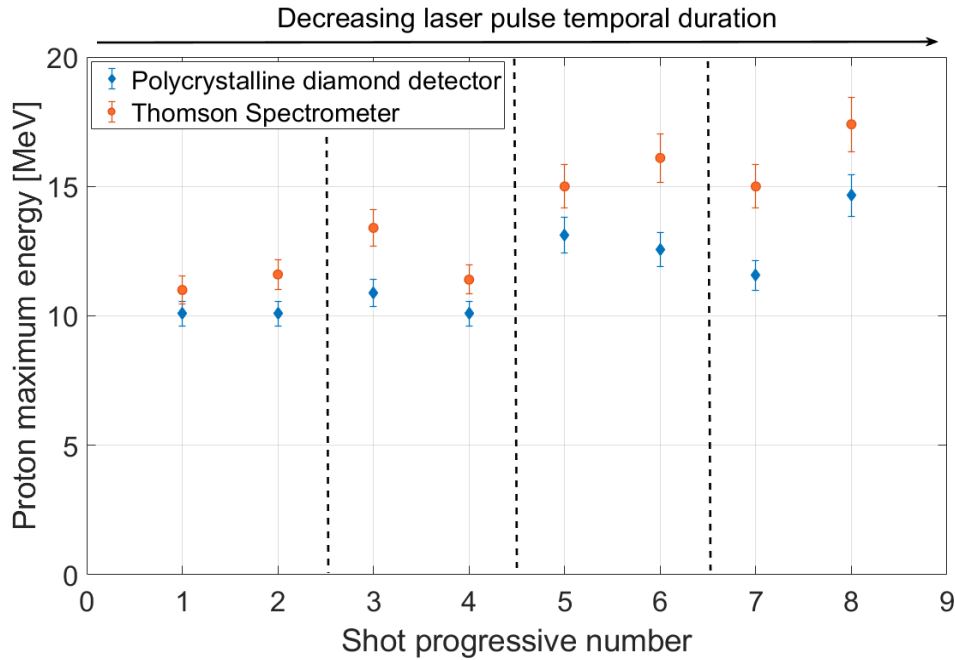

**Figure 3.1**. The maximum proton energies measured by the polycrystalline diamond detector are compared to those measured by means of a Thomson spectrometer placed on a similar angle. The shots were performed by irradiating a 3 µm aluminium target with a laser pulse energy ∼ 20 J and modifying the pulse duration every few shots, the dashed black lines points out the pulse length variation.

The information collected by the polycrystalline diamond allowed also to obtain a calibrated spectrum of the accelerated protons, by following the methodology fully described in *Salvadori*

– 6 –

*et al.* [16]. Nevertheless, as it is possible to observe in Figure 3.1, the proton cutoff energies are well above the 6 MeV limit. Therefore, to get the spectrum it was necessary to exploit equation (7) to include both the correction for the estimation of the number of protons at high energies and the filter attenuation factor. In Figure 3.2 the spectrum obtained from the data shown in Figure 2.1 is reported. The measured maximum proton energy during this shot is (18.5 ± 2.4) MeV.

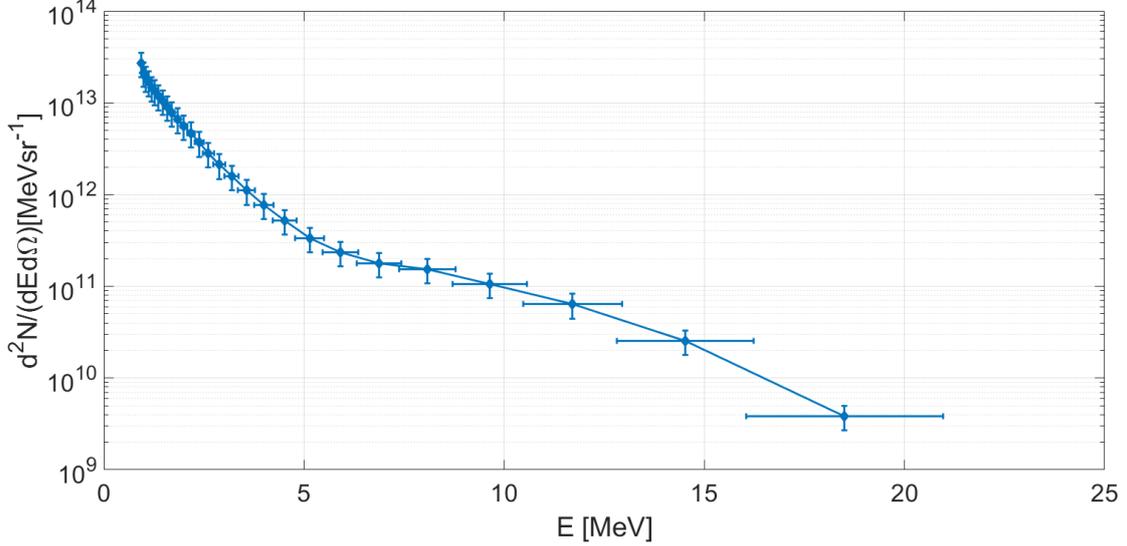

**Figure 3.2**. The proton spectrum retrieved from the raw signal shown in Figure 2.1. The correction factor $R(E_i)$ was applied to compute the number of particles for energies higher than 5.8 MeV.

## 4. Conclusion

Thanks to its features, the diagnostic system based on the large polycrystalline diamond detector can be employed for measurements at high sensitivity in environments strongly affected by electromagnetic pulses, offering the possibility to exploit the capability of many laser systems to work at high repetition rate. The time needed for data collection and storage is usually faster than the time needed for solid target alignment, thus it does not slow down the pace of the campaign. Moreover, the data analysis procedure, necessary to retrieve both the maximum energy of the accelerated protons and their spectrum, does not require complex signal post-processing. It is therefore well suitable to be implemented to run automatically providing all the useful information almost in real-time.

In this work we have shown that the system can be successfully used to monitor and characterize protons having energies much higher than 5.8 MeV, i.e., of protons that are fully stopped inside the detector. This is possible thanks to the high sensitivity of the diagnostic, obtained by the wide active surface, to the high signal to noise ratio and to the enhanced dynamic range offered by the developed methodology. These three factors, indeed, allow to detect with good accuracy even for the small amplitude signals generated by a low flux of energetic protons, and are thus capable to give effective description of the higher energy portion of the accelerated proton spectrum, where the number of particles decreases remarkably.

**Acknowledgments**

A gratefully thank to the team of the VEGA laser system at Centro de Laseres Pulsados for their expert support in performing the experiment.
The work has been carried out within the framework of the EUROfusion Consortium and has received funding from the Euratom research and training program 2014–2018 and 2019-2020 under grant agreement No. 633053. The views and opinions expressed herein do not necessarily reflect those of the European Commission.
The experiment at CLPU has received funding from LASERLAB-EUROPE V (Grant Agreement No. 871124, European Union Horizon 2020 research and innovation program) and from IMPULSE (Grant Agreement No. 871161, European Union Horizon 2020 research and innovation program). Support from Spanish Ministerio de Ciencia, Innovación y Universidades through the PALMA Grant No. FIS2016-81056-R, ICTS Equipment Grant No. EQC2018-005230-P, from LaserLab Europe IV Grant No. 654148, from Junta de Castilla y León Grants No. CLP087U16 and No. CLP263P20.